\documentstyle[prl,aps,twocolumn,epsfig]{revtex}

\begin{document}
\draft

\title
{
\bf Fractal Conductance Fluctuations in a Soft Wall Stadium
and a Sinai Billiard
}

\author
{A.S.~Sachrajda,$^{1}$ R.~Ketzmerick,$^{2}$ C.~Gould,$^{1,3}$ 
Y.~Feng,$^{1}$ P.J.~Kelly,$^{1}$ A.~Delage,$^{1}$ and Z.~Wasilewski,$^{1}$ }

\address {
$ ^1$Institute for Microstructural Sciences,
National Research Council of Canada, Ottawa, Canada K1A0R6\\
$ ^2$ Max-Planck-Institut f\"{u}r  Str\"{o}mungsforschung und Institut f\"{u}r
Nichtlineare Dynamik der
Universit\"{a}t G\"{o}ttingen,\\ Bunsenstr.~10, D-37073 G\"{o}ttingen, Germany \\
$^ 3$D\'{e}partement de Physique and CRPS, Universit\'{e} de Sherbrooke, Sherbrooke, Canada J1K 2R1
\medskip\\
\parbox{14cm}{\rm
Conductance fluctuations have been studied in a soft wall stadium
and a Sinai billiard
defined by electrostatic gates on a high mobility semiconductor heterojunction.
These reproducible magnetoconductance fluctuations are found to
be fractal confirming recent theoretical predictions of quantum signatures in
classically mixed (regular and chaotic) systems. The fractal character of the
fluctuations provides direct evidence for a hierarchical
phase space structure at the boundary between regular and chaotic motion.
}}
\maketitle
\narrowtext

 Ballistic geometries defined in two
dimensional electron gases (2DEGs) provide excellent
systems for studying quantum chaotic phenomena. In magnetic fields these systems
yield reproducible
 conductance fluctuations and weak localization effects analagous
to the universal conductance
 fluctuations and weak localization effects studied in disordered conductors.
In the ballistic case the phase
 coherent phenomena reflect the device geometry; not the random
positions of impurities. By studying these
 effects in both chaotic and non-chaotic geometries a valuable
insight has been obtained on how
 quantum information can be retrieved from classical chaotic dynamics. To date, in
virtually all of the
systems studied both experimentally and theoretically, a purely chaotic
system has been
assumed in which classical trajectories probe the phase space in a purely ergodic
fashion \cite{bluemel88,jalabert90,marcus92}
 before they exit. Escape from the ballistic cavity in such systems 
usually occurs exponentially fast.

 Real billiards, however, are typically not fully chaotic. They have a {\em mixed}
phase space, i.e., they contain both chaotic and regular regions.
Naively, one might expect that they are simply combinations of independent fully chaotic
and regular regions. But this is not the case. Mixed systems possess an important property which has drastic consequences for conductance fluctuations:
the chaotic part of phase space obeys a power law escape probability
 \cite{powerlaw}, in contrast to
the much faster exponential decay of fully chaotic systems. The power law
originates from chaotic trajectories which are 'trapped' close to an infinite
hierarchy of regular regions at the boundary between the regular and
chaotic motion (Fig.~3). This trapping is believed to be a
consequence of Cantori, which act as partial barriers for
transport and which surround the regular regions at all levels of the hierarchy
\cite{meiss85,geisel87}.
The phase of these long trapped trajectories is extremely sensitive to any
externally changed parameter. In a magnetic field, for example, they acquire a
phase factor
$\exp(2\pi i A B/\phi_0)$, which depends on the number $A B/\phi_0$
of magnetic flux quanta $\phi_0=h/e$ enclosed by the trajectory, ($A$
is the accumulated enclosed area).
The larger the value of $A$, the smaller the change in magnetic field required to modify the phase.  Therefore, in these 
systems, one expects to observe conductance fluctuations 
with much finer field scales \cite{lai92,rk96}. Semiclassical calculations 
\linebreak

\vspace*{0.7cm}
\begin{figure}[b]
\centering
\epsfig{figure=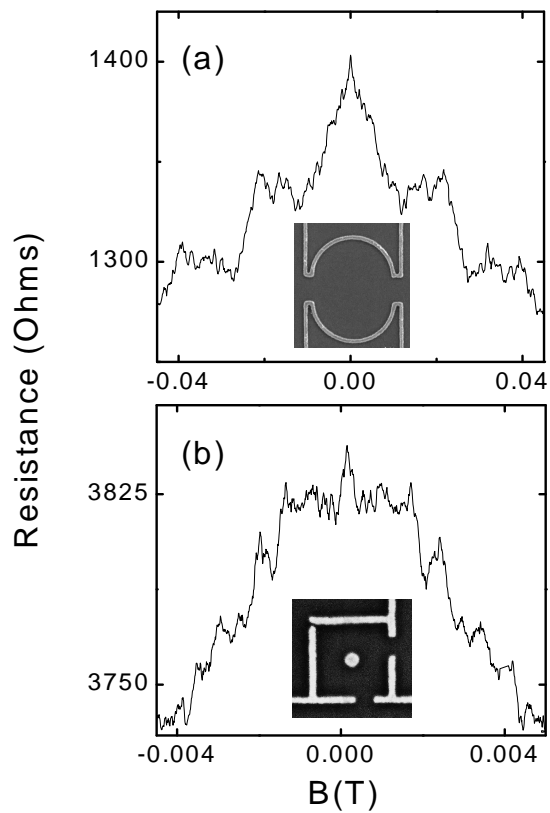,width=7.5cm}
\vspace*{0.2cm}
\caption{\footnotesize
Resistance vs.\ magnetic field (a) for the open stadium at a gate voltage of
-1.9V after illumination at 50mK and (b) for the Sinai Billiard at 50mK.
Fluctuations on both large and small scales can be seen for both devices. The two insets
are SEM micrographs of devices similar to the ones used. In the actual Sinai
billiard used for these measurements an insulator bridge connection (not shown)
was used to make contact to the center gate of the Sinai device.
}
\end{figure}

\noindent
similar to those in
Refs.~\cite{bluemel88,jalabert90,lai92}, have demonstrated
that a power-law distribution of enclosed areas larger than $A$,
\begin{equation}
P(A) \; \sim \;  A ^ {-\gamma} ,
\end{equation}
leads to {\em fractal} conductance fluctuations \cite{rk96}. Under these conditions the line
conductance versus magnetic field has the same properties as fractional
Brownian motion \cite{mandelbrot82}, which is self-affine in a statistical
sense and which is described by a fractal dimension, $D$, given by
\begin{equation}
D= 2 - \gamma /2 .
\label{fracdim}
\end{equation}
Thus the exponent $\gamma$, which describes a property of the
classical phase space, is related to the fractal dimension $D$
of a quantum coherent measurement.
 Similarly, the variance of conductance increments scale with small
 magnetic field increments as $<  (\Delta G) ^2 > \; \sim \; (\Delta
B)^{\gamma}$ . These results hold \cite{rk96} for $ \gamma \leq
 2$. Since  $\gamma > 1 $  must hold from more general arguments\cite
{meiss97} this
 restricts the fractal dimension to lie between 1 ($\gamma=2$) and 1.5
 ($\gamma=1$).
While the value of the classical exponent $\gamma$ is non-universal and is
 found numerically to
be sensitive to details of the geometry and the confining potential
\cite{rk96}, the
occurrence of power-law distributions is universal \cite{powerlaw}.
It is not feasable to test the existence of fractal conductance fluctuations
by numerical quantum calculations,
since a large number of modes are needed to adequately probe the
hierarchical classical phase space.
Indeed, because the semiclassical calculations of Ref.~\cite{rk96} are based on many
assumptions and approximations,
one may even speculate if fractal fluctuations exist at all.

The first experimental evidence for fractal behaviour was recently found in gold nanowires \cite{hegger}.
The variance analysis of the fluctuations showed a power law behavior over one
order of magnitude in magnetic field. While suggestive, this experiment suffered from two limitations,
i) it is well known that almost any smooth
function can be satisfactorily fitted by a power law for just one order of magnitude, and
ii) for the rectangular geometry of these gold nanowires one does not
expect a mixed phase space, unless fortuitous fluctuations in wire width
occasionally create such regions \cite{huckestein}.
Herein we report the first observation of fractal conductance
fluctuations over {\em two orders} of magnitude in magnetic field and
in {\it genuine} mixed phase space geometries.
The two classic models of fully chaotic systems, namely the stadium 
(two half circles connected by straight lines)
and the Sinai billiard (a square with a circular disk at its center)
are defined by electrostatic gates on a high mobility semiconductor
heterojunction. These systems are no longer fully chaotic due to soft-wall potentials, but have a mixed
phase space. This can be easily verified numerically (Fig.~3).
We find fractal conductance fluctuations as a function of magnetic field for large ranges of applied gate voltage (corresponding effectively to different
billiard sizes and potential forms).
 The existence of this new quantum signature of classically mixed systems is therefore   conclusively confirmed.
In addition, by choosing the stadium geometry, we have shown that
the famous experiment by Marcus et al.\ on the stadium 
\cite{marcus92}, which was performed to study the properties of fully chaotic
systems, gives qualitatively different results when carried out on todays
high mobility samples.

The two devices are shown as insets in Fig.~1. 
The stadium had a lithographic
radius of  $1.1 \mu m$. It was defined using metallic gates
on a high mobility AlGaAs/GaAs wafer
 (with a mobility of  $2.2 \times 10^6$ and $4.3 \times 10^6 cm^2/Vs$ and density of
$1.7 \times 10^{11}$ and $3.3 \times 10^{11} cm^{-2}$
 before and after illumination with a red LED). The
2DEG was 95nm below the surface. The device leads were made unusually wide (0.7
microns). This had the effect of
allowing most trajectories to rapidly exit the stadium with the exception of those
 trajectories which were trapped near the
 hierarchical phase space structure at the boundary between regular and
chaotic motion. This feature and the high mobility wafer used
 made this an optimal device for the observation of fractal conductance
fluctuations.
 The very high mobility wafer was necessary to achieve the required long phase
 coherence length, $l_\varphi$. The sub $mT$ features in the conductance
fluctuations and the narrow weak localization peaks ($<$ than  $150 \mu T$) confirmed that  $l_\varphi$ was
 indeed many times the perimeter of the stadium and Sinai billiard. Lithographic and fabrication details of the Sinai billard have been published elsewhere \cite{taylor97}. The measurements were made on
a dilution refrigerator using
 standard low power AC techniques at 50mK.

Measurements at higher temperatures ($\sim 4K$)  revealed small features
in the magnetoresistance related to
 the classical focusing of trajectories. As the temperature
was lowered both conductance fluctuations and a weak localization
 peak (around B=0) developed. In this paper we analyze the fluctuations which occur at magnetic fields for which the cyclotron diameter is
 larger than the stadium diameter. Features related to
 noise can be eliminated by comparing the $\pm B$ traces. 
 In Ref.~\cite{taylor97} it was observed for a soft wall Sinai billiard
that by rescaling magnetic field and resistance one finds quite similar
sequences of maxima and minima. This {\em non-statistical} self-affinity
remains unexplained so far and does not occur for the stadium,
 suggesting that this feature is
 characteristic of the Sinai billiard geometry.
We observe in contrast
conductance fluctuations on very different scales resembling the
{\em statistical} self-affinity of fractional Brownian motion for both geometries and for large ranges of gate voltage.

Experimental traces for both geometries are shown in Fig.~1. The results of a fractal analysis on these curves, shown in Fig.~2, gives fractal dimensions of $D=1.25$ and $D=1.30$.
The fluctuations are found to obey power-law scaling for over two orders of magnitude in magnetic field.
In all observable fractals this scaling behaviour is bounded by upper and lower cutoffs.
In our case the upper cutoff, in magnetic field, is determined by the smallest area for which the power law distribution holds
(which is close to the range of magnetic
fields experimentally studied).
The lower cutoff is determined by the minimum of two time scales:
i) the finite phase coherence time and
ii) the Heisenberg time $t_H=\hbar/\Delta E$, when the average level spacing
$\Delta E$ of the closed device is resolved and the semiclassical approximation
becomes unreliable.
The power law area distribution for trajectories staying in the device
longer than any of these times will not lead to fractal fluctuations.
The fractal dimension observed between these cutoffs is determined by using
a refined version of the box-counting algorithm. In the standard
box-counting algorithm one puts a grid of square boxes of size
$L \times L$ on the data (conductance vs.\ magnetic field) and counts
the number $N(L)$ of boxes through which the curve passes. Its dependence
on box size $L$, $N(L) \sim L^{-D}$, defines the fractal dimension $D$.
In this standard analysis the relative scale of
conductance and magnetic field is arbitrary so that for a finite data set
the resulting fractal dimension depends on the aspect ratio of
the plot.
To overcome this limitation we applied a
modified version of the algorithm: This
divides the magnetic field range in length intervals  $\Delta B$ and
determines $N(\Delta B)$ as the difference of maximum and minimum
conductance in each interval, summed over all intervals, and divided by
$\Delta B$. This corresponds in the standard algorithm to taking rectangular
boxes with infinitely small size in the conductance direction.
The inset of Fig.~2 
shows the analysis of the variance, which gives $\gamma=1.38$ 
and $\gamma=1.30$, for stadium and Sinai billiard respectively,
confirming the above fractal analysis \cite{spectrum}.

\begin{figure}[b]
\centering
\epsfig{figure=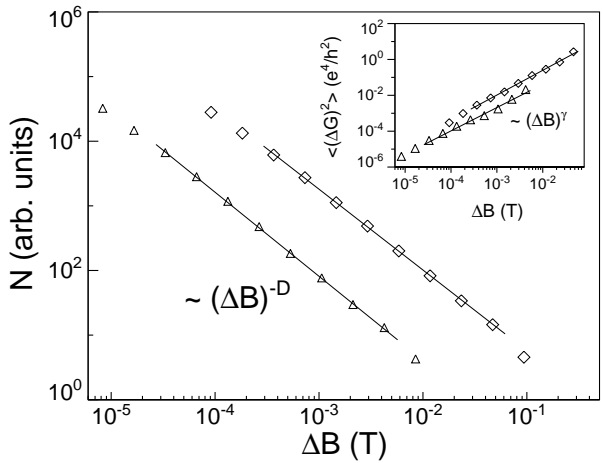,width=8.5cm}
\vspace*{0.3cm}
\caption{\footnotesize
Fractal analysis for the stadium (diamonds) and Sinai device (triangles)
data
shown in Fig.~1 using a modified box-counting algorithm.
The number of boxes $N$ follows power laws $(\Delta B)^{-D}$ for two
ordes of magnitude in magnetic field scale, giving fractal dimensions
$D=1.25$ and $D=1.30$, resp.
The inset shows the variance analysis giving $\gamma=1.38$ and
$\gamma=1.30$, resp., in reasonable agreement with Eq.~(2).
}
\end{figure}

\begin{figure}[b]
\centering
\epsfig{figure=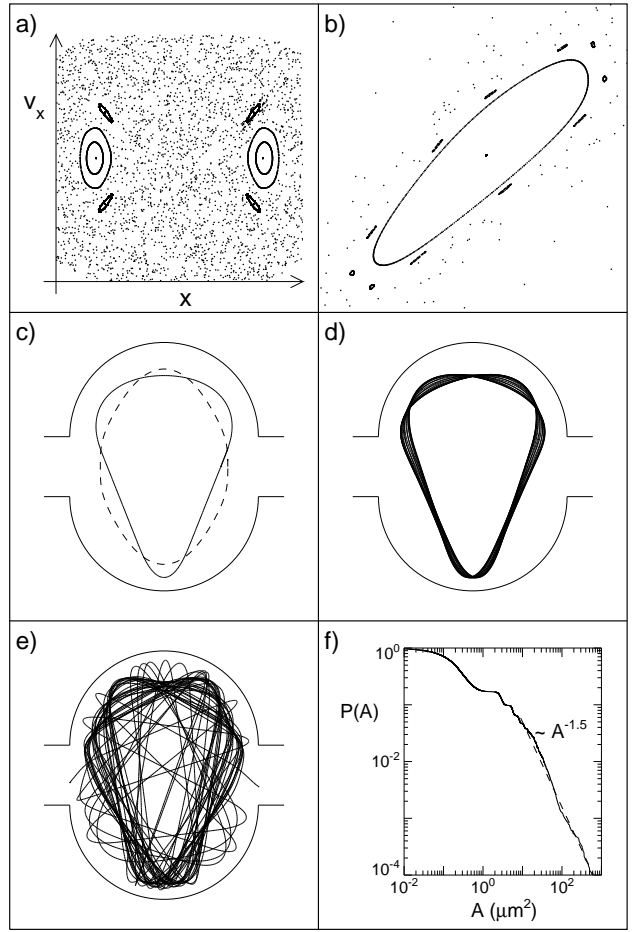,width=8.2cm}
\vspace*{0.3cm}
\caption{\footnotesize
(a) Poincar\'e surface of section of the stadium of Fig.~1 (with
$0.1 \mu m$ depletion length and a $0.4 \mu m$ wide parabolic wall).
It shows $v_x$ vs.\ $x$ at every trajectory intersection  with the
horizontal symmetry line whenever $v_y > 0$ holds, for one chaotic and 6
regular
trajectories.
(b) Enlargement of the small island in (a) showing higher order islands.
(c) Periodic trajectories corresponding to the center of the large island in
(a) (dashed line)
and the center of the island in (b) (solid line).
(d) Quasiperiodic trajectory corresponding to the island in (b).
(e) Chaotic trajectory being trapped in the stadium close to the regular
trajectory of (d).
(f) Integrated area distribution displaying a power law for large enclosed
areas.
}
\end{figure}

We now show that the stadium, a classic model
for fully chaotic systems, has a mixed classical phase space when achieved electrostatically due to the soft-wall potential experienced by the electrons. We model this potential by a depletion region of 100nm
at the edge of the device followed by a 400nm parabolic region and a flat
potential beyond that \cite{potential}.
 The mixed phase space is revealed in a Poincar\'e section analysis (Fig.~3).
Its hierarchical structure gives rise to long trapped trajectories
as well as a power-law area distribution.
The effective Planck's constant $\hbar_{\rm eff} =\hbar / \int p {\rm d}q = (L
\sqrt{2 \pi n})^{-1}$
for a trajectory along the circumference of the stadium is $\hbar_{\rm eff} =
8.6 \times 10^{-4}$.
All of this confirms the applicability of the results of
Ref.~\cite{rk96} to the present experiment.
Similar conclusions apply to the soft-wall Sinai billiard.

Direct Simulation is the only technique available
to predict the exact value of
the classical power-law exponent and the corresponding fractal dimension
 for a particular device.
 The values of these exponents depend critically on the exact 
form of the potential \cite{rk96}.
But even if the electrostatic potential due to the gates is known,
it would be modified in a real device by the presence of disorder.
In high-mobility samples, however, the disorder potential is
weak and smooth
and so does not change the qualitative character of the classical phase space,
i.e., it
remains a mixed phase space. The exponent $\gamma$ of the power law area
distribution of trapped
chaotic trajectories on the other hand, might well be changed by the specific
disorder configuration
inside the cavity. Thus one cannot hope for a quantitative comparison of the
exponent $\gamma$
deduced from classical simulations without disorder with the fractal dimension
$D$ of the
conductance fluctuations of a real device containing disorder.
Disorder outside the cavity may in principle also change the conductance
fluctuations but it cannot
influence the trapping of chaotic
trajectories inside the cavity and thus it will have no significance for the observed
fractal dimension.
As a function of the applied gate voltage, the exponents of the classical power law distribution may fluctuate for large times\cite{laiding92} but  
 for times smaller than the phase
coherence time the power law exponents have been found to be much
more stable \cite{rk96}. This feature is also
 reproduced in our experiment. Figure 4 reveals that there exists a monotonic
relationship between the fractal
 dimension and the gate voltage applied to the stadium device. The
 fractal dimension tends towards 1 as the confinement is reduced.

\begin{figure}[b]
\centering
\epsfig{figure=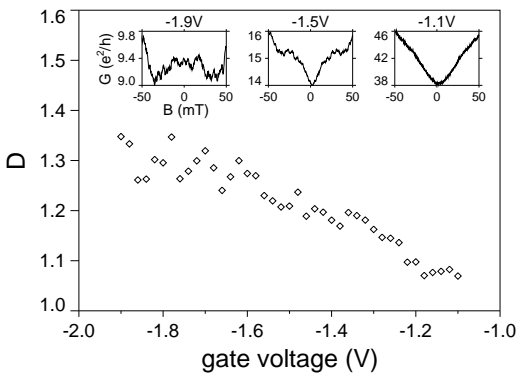,width=7.0cm}
\vspace*{0.2cm}
\caption{\footnotesize
Fractal dimension of conductance fluctuations vs.\ gate voltage for the
stadium
before illumination. Error bars of the fractal analysis
are typically $\pm 0.05$. The insets show data for three gate voltages.
}
\end{figure}

	We have reported the first observation of fractal conductance
fluctuations in semiconductor billiard devices  \cite{micolich}.
The observation confirms a recent theoretical prediction that conductance
fluctuations in mixed phase space systems are statistical self-affine
and can be described by a fractal dimension.
The origin of the behaviour is a power law distribution of areas enclosed
by chaotic trajectories, which results from the hierarchical structure
of phase space at the boundary of regular and chaotic motion.
We have shown that in real devices the classification
into 'chaotic' and 'regular' geometries is incomplete at best.
Our findings confirm the important role that soft-wall potentials play
in nanostructures. We have also observed a dependence of the fractal dimension of conductance
fluctuations on the device parameters.

We would like to acknowledge the assistance of P.~Zawadzki with data acquisition, P.T.~Coleridge for useful discussions and R.~Newbury and R.~Taylor for assistance with the Sinai device measurements.


\begin{references}
\bibitem{spectrum} Similarly, one may study the power spectrum, as was done in
Ref.~\cite{marcus92},
for which we predict a power law $S(\omega) \sim \omega^{-(\gamma+1)}$ for a
mixed phase space.
Numerically, one needs to average $S(\omega)$ over different magnetic field
regions, making this
analysis inappropriate for the short magnetic field traces of this experiment.
\bibitem{potential} In the entrance and exit regions we add the potential
coming from the upper and lower metallic gate.
Other parameters for the potential give qualitatively similar results.
\end{references}
\end{document}